\documentclass[nonatbib]{article}

\usepackage[final]{nips}

\usepackage[utf8]{inputenc} 
\usepackage[T1]{fontenc}    
\usepackage{hyperref}       
\usepackage{url}            
\usepackage{booktabs}       
\usepackage{amsfonts}       
\usepackage{nicefrac}       
\usepackage{microtype}      
\usepackage{graphicx}
\usepackage{multirow}

\usepackage{amssymb}
\usepackage{pifont}

\usepackage{algpseudocode}

\title{Mixed-initiative Query Rewriting in Conversational Passage Retrieval}

\author{%
  Dayu Yang \\
  University of Delaware\\
  \texttt{dayu@udel.edu} \\
   \And
   Yue Zhang \\
   University of Delaware\\
   \texttt{zhangyue@udel.edu} \\
   \AND
   Hui Fang \\
   University of Delaware\\
   \texttt{hfang@udel.edu} \\
}

\begin{document}

\maketitle

\begin{abstract}

In this paper, we report our methods and experiments for the TREC Conversational Assistance Track (CAsT) 2022. In this work, we aim to reproduce multi-stage retrieval pipelines and explore one of the potential benefits of involving mixed-initiative interaction in conversational passage retrieval scenarios: reformulating raw queries. Before the first ranking stage of a multi-stage retrieval pipeline, we propose a mixed-initiative query rewriting module, which achieves query rewriting based on the mixed-initiative interaction between the users and the system, as the replacement for the neural rewriting method. Specifically, we design an algorithm to generate appropriate questions related to the ambiguities in raw queries, and another algorithm to reformulate raw queries by parsing users' feedback and incorporating it into the raw query. For the first ranking stage of our multi-stage pipelines, we adopt a sparse ranking function: BM25, and a dense retrieval method: TCT-ColBERT. For the second-ranking step, we adopt a pointwise reranker: MonoT5, and a pairwise reranker: DuoT5. Experiments on both TREC CAsT 2021 and TREC CAsT 2022 datasets show the effectiveness of our mixed-initiative-based query rewriting (or query reformulation) method on improving retrieval performance compared with two popular reformulators: a neural reformulator: CANARD-T5 and a rule-based reformulator: historical query reformulator(HQE).

\end{abstract}

\section{Introduction}
The TREC Conversational Assistance Track (CAsT) is a task to facilitate the study of conversational information systems, which are information systems that adopt a conversational modality to enable conversational exchanges between the system and its users. The main objective of conversational information seeking is to satisfy users' information needs in an evolutionary fashion, which is formalized or expressed through conversation turns. It can be beneficial for many information retrieval tasks, such as sophisticated information searching, exploratory information collecting, multi-turn retrieval task completion, and recommendation. Although conversation can also exhibit other types of interactions with different characteristics and modalities, such as clicks, multi-choice selections, and other forms of feedback~\cite{zamani2022conversational}, we mainly focus on natural language conversation in the Text REtrieval Conference(TREC) Conversational Assistance Track(CAsT). Specifically, following the problem settings of TREC CAsT, a user can initialize an open-domain information request to the system, and the system is expected to retrieve relevant passages from a gigantic corpus. During the conversation, the user is free to continue on the previous topic, provide feedback on the previously retrieved passage, or shift from one topic to another.

The overall approach of us is a multi-stage retrieval architecture that contains four main stages: the query rewriting stage, the first-ranking stage, and the second-ranking stage with an affiliated stage called fusion. In addition to adopting existing query reformulation methods: CANARD-T5 and HQE\cite{lin2021multi}, to enable the mixed-initiative interaction and make the interaction helpful to the query rewriting task, we design an algorithm to generate questions seeking clarification of three types of ambiguities in the raw queries: incomplete, reference, and descriptive. After the question is generated, it will be sent to the user. Once the answer from the user is received, the answer will be parsed by another algorithm. The new clarification information will be combined with the raw query to formulate the reformulated query.

\section{Method}

\subsection{Multi-stage Retrieval Pipeline}
\begin{figure}[h]
    \centering
    \includegraphics[scale=0.4]{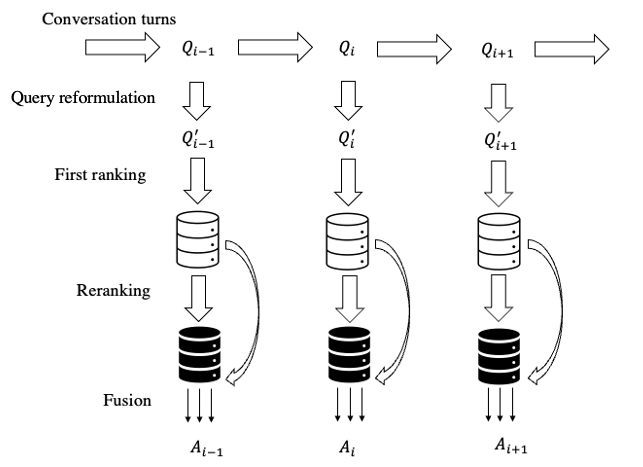}
    \caption{Demonstration of our multi-stage conversational text retrieval pipeline for creating non-mixed-initiative runs}
    \label{fig:multi_stage}
\end{figure}

First, in order to achieve state-of-the-art retrieval performance, we construct an efficient and effective multi-stage retrieval pipeline. Specifically, we use a four-stage cascade structure. The first stage will be the query rewriting stage, where we implement two popular query rewriting methods: CANARD-T5 rewriter and HQE~\cite{lin2021multi} to eliminate ambiguities in the raw utterances. The pipeline we build can be illustrated in Figure~\ref{fig:multi_stage}.

To improve the efficiency of the multi-stage pipeline, instead of reformulating the query when we run the pipeline, we first implement query rewritings on all the queries and store the reformulated queries for later usage. This reduces the intensity of communication between the CPU and GPU and avoids unnecessary repeated query rewriting work when trying different ranking functions. For the CANARD-T5 rewriter~\cite{lin2021multi} which we use for query rewriting, instead of using the parameters, we fine-tune it on TREC 2019 and 2020 data to further improve the reformulation on retrieval. The experiment on the TREC CAsT 2021 dataset shows our fine-tuned T5 rewriter outperforms the T5 with original weights.

After the ambiguities are resolved by the query rewriting stage, the first ranking stage generates the initial ranked document list and delivers it to the re-ranking stage. there will be $\gamma_1$ numbers of documents to be retrieved as the candidates for the following stage. We apply multiple ranking methods based on sparse methods and dense methods to overcome the limitation of the lexical and semantic matching capability of a single ranking function.  Each $stage_i$ takes $\gamma_{i-1}$ numbers of documents to ranking and outcome $\gamma_i$ numbers of documents to the subsequent stage, where $\gamma_{i-1} \le \gamma_i$. Specifically, we use BM25 ranking function as the sparse retrieval method we used in the first ranking stage. For the dense ranking in the first ranking stage, we use TCT-ColBERT~\cite{lin2020distilling} to independently encode queries and the documents and formulate the dense representations of documents and queries. After the first ranking stages are finished, we use a pointwise re-ranker MonoT5~\cite{nogueira2020document} and a pairwise re-ranker DuoT5~\cite{pradeep2021expando} to re-rank the relevant documents (passages) that are outputted from the first ranking stage. Finally, we apply reciprocal rank fusion\footnote{We do not use early fusion. The fusion step is always employed after the second-stage ranking}, which is efficient for combining the ranked lists obtained from re-ranking. (If an early fusion is applied, the fusion stage will behave before re-ranking and after the first ranking. After fusion, the final ranked document list will output as result. Figure~\ref{fig:multi_stage} shows our multi-stage conversational text retrieval pipeline for creating non-mixed-initiative runs. 

\subsection{Mixed-initiative Query Rewriting}

We observe that, for the TREC CAsT 2021 dataset, in some cases, certain ambiguities are not successfully clarified by the CANARD-T5 reformulators. Since CANARD-T5 is a generative model that samples the reformulated queries from high-dimensional distributions. It is hard for us to explicitly explore why sometimes CANARD-T5 fails to correctly clarify ambiguities. However, what we can observe is that: the automatic reformulator performs well on certain queries but not well on the rest. Therefore, with the introduction of the mixed-initiative task in the TREC CAsT 2022, we intend to explore the potential of clarifying the ambiguity with the help of users. In order to generate the question, we designed an algorithm to identify the ambiguity a raw query has and formulate the corresponding question. 

\begin{figure}
    \centering
    \includegraphics[scale=0.3]{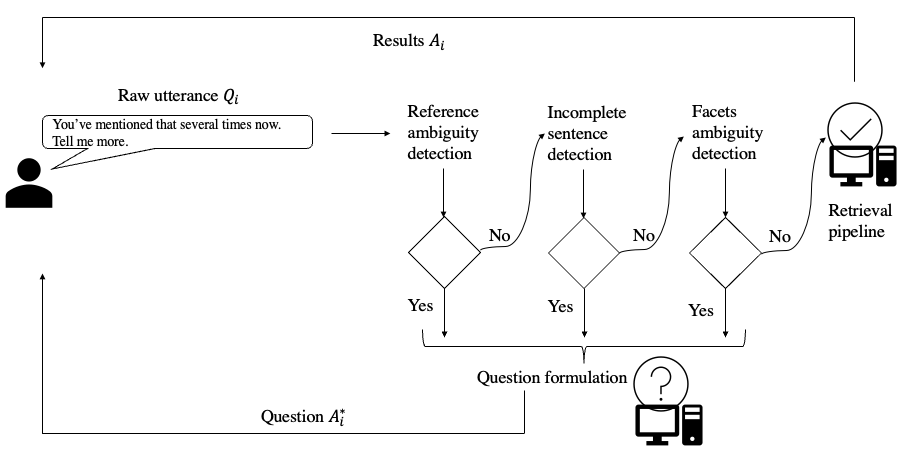}
    \caption{A workflow for asking clarifying questions in an open-domain conversational search system}
    \label{fig:clarify}
\end{figure}

\vspace{0.2cm}

Specifically, we design three questions that correspond to three types of ambiguities: references, descriptive, and incomplete sentences. When the algorithm detects an ambiguity, the system will generate a corresponding question from the template and send it to the user for further clarification. \textit{Incomplete} ambiguity is defined as the raw query does not have any nouns. For example, the user may ask ``How's that?" after the previous retrieval turns. \textit{Reference} ambiguity is defined as the algorithm finding at least one pronoun in the raw query(and it is not at the beginning of the raw query). \textit{Descriptive} ambiguity means the noun with a high BM25 score in the raw query does not have any descriptive information following it. For example, the raw query could be ``What kind of innovation do we have?" which is missing the description of the noun ``innovation". The descriptive information of ``innovation" can be ``innovation of US banks". 

Since only one interaction is allowed for each raw query, if multiple types of ambiguities are detected, the algorithm will follow the priority order: incomplete $>$ reference $>$ descriptive to generate questions. Here are some more concrete examples from the TREC CAsT 2022 dataset, in \textit{utterance 3-1, turn 142}, the raw query is \textit{You’ve mentioned that several times now. Tell me more.}. The algorithm will first identify if this sentence is a complete query. In this case, the raw query is a complete query, so the algorithm will move on to detect the reference ambiguity. Since there is a pronoun ``that" in the raw query, that means the raw query has a reference ambiguity. Then the algorithm will extract the pronoun ``that" and ask the user, \textit{What does ``that" refer to in your raw query?}. If there is no reference ambiguity found by the algorithm, it will continue to try to identify whether raw utterances have descriptive ambiguities. Taking utterances 3-3, turn 132 as an example, the raw utterance is \textit{How did other parties respond?}. The algorithm detected that an important word, \textit{parties} does not have any descriptive information in the raw utterance. Therefore, the system will ask the user to specify the word \textit{parties}. In this case, the answer received by the system is \textit{parties of the Paris Agreement}. 

After the answer of obtained, a reformulating algorithm will add the newly-obtained information from the answer to the original query. For an original query with the reference ambiguity, we expect the user's answer to be the entity a pronoun is referring to. So the pronoun will be directly replaced by the answer. For an incomplete original query, we expect the user's answer to be the complete query. So the entire original query will be replaced by the answer. For an original query with descriptive ambiguity, the algorithm will append the descriptive information to the corresponding noun or verb. In some cases, the user may refuse to answer the question, the answer will be ``I don't know" if the user refuse to answer in TREC CAsT 2022. The algorithm will keep the original query intact if it meets the answer ``I don't know".

In summary, the algorithm can enable mixed-initiative interaction with users while resolving the ambiguities that existing generative and classification query rewriting methods have difficulty resolving. The overall workflow of our mixed-initiative algorithm is shown in Figure~\ref{fig:clarify}. 

\begin{figure}[h]
    \centering
    \includegraphics[scale=0.4]{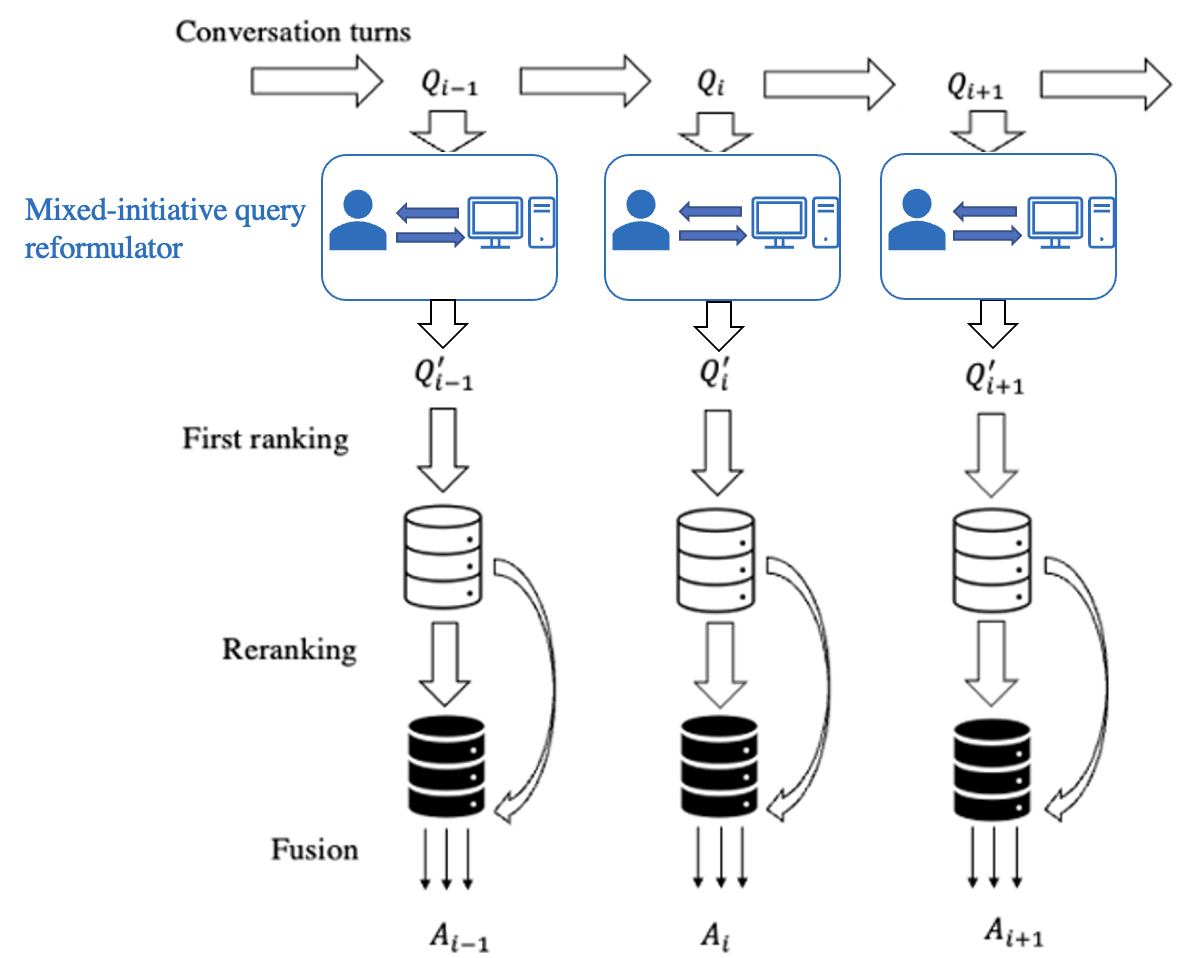}
    \caption{Demonstration of our multi-stage conversational text retrieval pipeline for creating mixed-initiative runs}
    \label{fig:mix-multi}
\end{figure}

Overall, the original automatic query rewriting steps of the multi-stage retrieval pipeline introduced in Figure~\ref{fig:multi_stage} are replaced with the mixed-initiative query rewriting module, which means the change is made only in the ``query rewriting" stage. The multi-stage conversational text retrieval pipeline we used for creating mixed-initiative runs for TREC CAsT 2022 can be illustrated in Figure~\ref{fig:mix-multi}.

\vspace{0.2cm}

\section{Experimental Setup}

\subsection{CAsT Datasets}

Since the qrel file of the TREC CAsT 2022 dataset is not available when we participate in the CAsT track. We finish experiments about hyperparameter tuning on the TREC CAsT 2021 dataset.

\subsection{Query Rewriting Setup}

The original T5 rewriter has trained on CANARD~\cite{lin2021multi} dataset. Although many experiments from the runs of TREC CAsT 2021 show the knowledge T5 learns from CANARD transfers well to the TREC CAsT dataset, we still want to figure out \textit{if it is beneficial to fine-tune the T5 rewriter on the previous TREC CAsT datasets: 2019 and 2020?} Therefore, we start with the weights that we borrowed from~\cite{elgohary2019can} and fine-tune the T5 on structured TREC 2019 and 2020 data. Table~\ref{table:finetuning} shows: by using the query reformulated by the fine-tuned T5 rewriter, the retrieval performance on the first ranking stage can surpass the original T5 rewriter.

\begin{table}[]
\centering
\begin{tabular}{c|cccc}
\hline
Run Name                 & Recall@1000 & Recall@500 & MAP@2000 & NDCG@3 \\ \hline
The original T5 rewriter & 0.5873      & 0.5502     & 0.1378   & 0.2335 \\ \hline
Fine-tuned T5 rewriter   & \textbf{0.6365}      & \textbf{0.5892}     & \textbf{0.1484}   & \textbf{0.2525} \\ \hline
\end{tabular}
\vspace{0.2cm}
\caption{The retrieval performance on TREC CAsT 2021 dataset of the reformulated queries reformulated by the original T5 rewriter from~\cite{elgohary2019can} and our fine-tuned one.}
\label{table:finetuning}
\end{table}

\vspace{0.2cm}

For the default setting of a T5 rewriter, it will consider all the canonical passages $A_{<i}: A_1, A_2, …,A_{i-1}$ before the focal query $Q_i$. However, when applying it to the TREC CAsT dataset, the canonical passage is usually much longer than the canonical passage in the CANARD dataset. The excessive length brings up the issue that the total length of the input of T5 will sometimes surpass the maximum length of the token input of T5: 512. And T5 will cut off all the tokens after the $512_{th}$ token. This makes some latest users' utterances may be abandoned by the T5's tokenizer. However, the intuition is that the user's information needs are more likely to be contained in users' utterance $Q_{<i}: Q_1, Q_2, … ,Q_{i-1}$ instead of the canonical passages $A_{<i}$. Another issue is that the long canonical passages may bring T5 extra challenges to locate useful information from the context that contains the information about users' implicit information needs. Therefore, we think concatenating all previous canonical passages may harm the retrieval performance. The results of experiments on the TREC CAsT 2021 dataset are shown in Table~\ref{table:num_canonical}. As we can see, the best-performing reformulated queries only take the most recent canonical passage into consideration.

\begin{table}[h]
\centering
\begin{tabular}{c|cccc}
\hline
No. Canonical Psg* & Recall@500 & MAP@500 & NDCG@500 & NDCG@3 \\ \hline
0                                             & 0.5330      & 0.1237  & 0.3384   & 0.2180  \\ \hline
1                                             & \textbf{0.5459}     & \textbf{0.1356}  & \textbf{0.3591}   & \textbf{0.2474} \\ \hline
2                                             & 0.4688     & 0.1111  & 0.3012   & 0.2102 \\ \hline
3                                             & 0.4688     & 0.1111  & 0.3012   & 0.2102 \\ \hline
\end{tabular}
\vspace{0.2cm}
\caption{The retrieval performances on TREC CAsT 2021 dataset of the reformulated queries considering the different numbers of canonical passages(*No. Canonical Psg = Number of canonical passages to consider in the T5 rewriter)}
\label{table:num_canonical}
\end{table}

\vspace{0.2cm}

Another experiment we did was to explore ``\textit{if it is beneficial to consider not only the most probable output of the generative reformulator but other outputs}?" The intuition behind this experiment is that the target of a generative model is to generate the sentence with the highest probabilities instead of a sentence with more information that represents users' information needs. By our observation, we find, sometimes, the probability of generating the reformulated sentence: ``\textit{What is the price of the bike}?" is larger than the probability of ``\textit{What is the price of the sport bike of Trek}?" Therefore, we design an experiment to fuse top-probable sentences from the generative reformulator. The results are shown in Table~\ref{table:num_sentence}. We can see that the recall can be largely improved if we consider multiple top-probable outputs and fuse them at the end of the first ranking stage. The results indicate that it may be beneficial to consider multiple generated sentences with the largest probabilities from a generative reformulator such as T5 rewriter.

\begin{table}[]
\centering
\begin{tabular}{c|cccc}
\hline
No. sentences to fusion & Recall@500 & MAP@500 & NDCG@500 & NDCG@3 \\ \hline
1                           & 0.4854     & 0.1294  & 0.3252   & 0.2429 \\ \hline
3                           & 0.5663     & 0.1376  & 0.3662   & 0.2466 \\ \hline
5                           & 0.5793     & 0.1431  & 0.3758   & 0.2607 \\ \hline
7                           & 0.5845     & \textbf{0.1503}  & \textbf{0.3833}   & \textbf{0.2722} \\ \hline
10                          & \textbf{0.6037}     & 0.1423  & 0.3804   & 0.2587 \\ \hline
\end{tabular}
\vspace{0.2cm}
\caption{The retrieval performances on TREC CAsT 2021 dataset of runs fusing different number of top-probable sentences in the first ranking stage (using BM25 as the ranking function)}
\label{table:num_sentence}
\end{table}

Historical query rewriting(HQE) is a method that uses BM25 scores to identify if a word in the previous passage retrieval log is important or not. The words that are classified as important will be appended at the end of the raw query. We use the same BM25 threshold setting with the paper introducing HQE\cite{lin2021multi}.

\subsection{Tuning BM25 Parameters}

For sparse ranking functions, we use the BM25 ranking function with the following parameter setting: k1=1.24, b=0.9 after hyperparameter tuning on the TREC CAsT 2021 dataset. Table~\ref{table:bm25_tuning} shows the improvement in retrieval performance using the aforementioned BM25 parameters compared with the default settings in the TREC CAsT 2021 dataset. For both sparse retrieval and dense retrieval in the first ranking stage, our pipeline will only return the top 2000 ranked passages to improve efficiency. Also, during the first retrieval, instead of waiting for the retrieval process to finish for a single query, we delay all the retrieval processes for many reformulated queries and process them together to take full advantage of the multiprocessing capability of our CPU.

\begin{table}[]
\centering
\begin{tabular}{c|ccc}
\hline
                                & Recall@500 & Recall@1000 & Recall@3000 \\ \hline
A non-tuning run                & 0.6664     & 0.7121      & 0.7790      \\ \hline
The best run(k1=1.24,   b=0.9). & \textbf{0.6730}     & \textbf{0.7518}      & \textbf{0.8101}    \\ \hline 
\end{tabular}
\vspace{0.2cm}
\caption{The improvement in retrieval performance using the aforementioned BM25 parameters compared with the default settings in the TREC CAsT 2021 dataset}
\label{table:bm25_tuning}
\end{table}

\subsection{Re-ranking Setup}

After the first ranking stage, we implement MonoT5 as the pointwise re-ranker and DuoT5 as the pairwise re-ranker. Due to the expensive computational requirement that come with the re-ranking process and re-rankers require online computation. We only use MonoT5 to re-rank the first 1000 documents and use DuoT5 to re-rank the first 200 documents from MonoT5.

\subsection{Fusion Setup}

After the re-ranking stage, we have a total of six re-ranking runs since we have three different versions of reformulated queries: G0, G1, and Hqe, and two different first ranking methods: sparse and dense ranking methods, where``G" stands for generative neural reformulator. In our case, we use the T5 rewriter specifically.  We consider both ``G0" and ``G1", where ``G0" stands for the most probable output and ``G1" stands for the second most probable output. Although we observe from Table~\ref{table:num_sentence} that fusing more runs using different output of the generative reformulator could be beneficial to the retrieval performance. Due to the time restriction and for the simplicity of our method, we do not fuse runs using top-probable sentences generated by the reformulator for creating our submitted runs. The following chapter will include a detailed description of the four submitted runs and how we finished the fusion step.

\section{Submitted Runs}

\begin{table}[]
\centering
\begin{tabular}{c|c|c|ccc}
\hline
Method           & Dataset               & MI  & Recall@1000 & MAP@1000 & NDCG@3 \\ \hline
udinfo\_best2021 & \multirow{2}{*}{CAsT 2021} & no  & \textbf{0.949}       & \textbf{0.287}    & \textbf{0.246}  \\ \cline{3-3}
udinfo\_onlyd    &                       & no  & 0.904       & 0.286    & 0.240   \\ \hline
udinfo\_best2021 & \multirow{4}{*}{CAsT 2022} & no  & 0.681       & 0.181    & 0.325  \\ \cline{3-3}
udinfo\_onlyd    &                       & no  & 0.651       & 0.178    & 0.348  \\ \cline{1-1} \cline{3-3} 
udinfo\_best2021\_mi &                       & yes & \textbf{0.771}       & \textbf{0.246}    & \textbf{0.452}  \\ \cline{3-3}
udinfo\_onlyd\_mi    &                       & yes & 0.729       & 0.243    & 0.450   \\ \hline
\end{tabular}
\vspace{0.2cm}
\label{table:result}
\caption{The retrieval performance on the TREC CAsT 2021 and 2022 datasets of all submitted runs. MI stands for ``mixed-initiative".}
\end{table}



For TREC CAsT 2022 participation, our team finally submitted two automatic runs: UDInfo-best2021, UDInfo-onlyd and two mixed-initiative runs: UDInfo-mi-b2021, UDInfo-onlyd-mi. Since we cannot obtain user feedback for the TREC CAsT 2021 dataset, we only report the two pipelines using CANARD-T5 or HQE query reformulators. The details of how to create runs in~\ref{table:result} are described later.

What we can observe from~\ref{table:result} is that: first, although the two methods we used for creating runs have a higher Recall@1000 on the TREC 2021 dataset compared with the TREC 2022 dataset, we obtain higher NDCG@3 on the TREC 2022 dataset. This could indicate that the ranking methods we used in the first ranking stage, which mainly focused on increasing Recall, can handle the TREC 2021 dataset better than the TREC 2022. On the contrary, the ranking methods we used for the second-ranking stage, which mainly focuses on increasing NDCG, can handle the TREC 2022 dataset better than the 2021 one. Secondly, the methods using the mixed-initiative query rewriting module achieve much higher retrieval performance compared with the runs using CANARD-T5 and HQE, which indicates that incorporating mixed-initiative interaction into conversational passage retrieval systems has the potential to improve retrieval performance\footnote{Since some of the original answers we received have unexpected bad quality(for example, many answers are ``This question is not related to my search."), we manually replace those bad-quality answers by mimicking the behavior of a user. The answer file, which includes the answers we used for query rewriting and our generated questions, can be found in \href{https://drive.google.com/file/d/1qO0xvMNGdoMJyD_43CQFqLPygI4QuA7_/view?usp=sharing}{this link}.}.

\begin{itemize}
    \item Run \#1 (UDInfo-best2021): reciprocal fusion of three top-ranked methods on NDCG@3 in the TREC CAsT 2021 dataset(The reason for only fuse top three runs is because fusing more runs can only harm the performance by experiment on the TREC CAsT 2021 dataset), which are:
    \begin{itemize}
        \item Using ``G1" as the query rewriting method; sparse first ranking stage; Pointwise and Pairwise re-ranking.
        \item Using ``G0" as the query rewriting method; dense first ranking stage; Pointwise and Pairwise re-ranking.
        \item Using ``Hqe" as the query rewriting method; dense first ranking stage; Only re-ranking on Pointwise method.
    \end{itemize}
    \item Run \#2 (UDInfo-mi-b2021): using the clarification answers from users after the system proactively elicits; other remains the same as Run \#1.
    \item Run \#3 (UDInfo-onlyd): reciprocal fusion of all three dense methods, which are:
    \begin{itemize}
        \item Using ``G1" as the query rewriting method; dense first ranking stage; Pointwise and Pairwise re-ranking.
        \item Using ``G0" as the query rewriting method; dense first ranking stage; Pointwise and Pairwise re-ranking.
        \item Using ``Hqe" as the query rewriting method; dense first ranking stage; Pointwise and Pairwise re-ranking.
    \end{itemize}
    \item Run \#4 (UDInfo-onlyd-mi): using the clarification answers from users after the system proactively elicits; other remains the same as Run \#3.
\end{itemize}



\section{Conclusion}

In this paper, we introduced our multi-stage retrieval pipeline that can tackle conversational search tasks. Our pipeline consists of four stages: query rewriting, first ranking, re-ranking, and fusion. In addition to the multi-stage retrieval pipeline, we also introduced our implementation of mixed-initiative interaction on query rewriting, where we design an algorithm to generate questions and seek answers from users to explicitly resolve the ambiguities in the raw queries. In the future, we will explore more methods that can enable mixed-initiative interactions, which can possibly benefit retrieval performance in conversational search.

\newpage

\bibliographystyle{plain}
\bibliography{ref.bib}

\end{document}